# Small metal artifact detection and inpainting in cardiac CT images

Trevor McKeown[1], H. Michael Gach[3,4,5], Yao Hao[3], Hongyu An[4,5], Clifford G. Robinson[3], Phillip S. Cuculich[6], Deshan Yang[2]

[1]Medical Physics Program, Duke University
[2]Department of Radiation Oncology, Duke University
[3]Department of Radiation Oncology, School of Medicine, Washington University in Saint Louis
[4]Department of Radiology, School of Medicine, Washington University in Saint Louis
[5]Department of Biomedical Engineering, Washington University in Saint Louis
[6]Department of Cardiology, School of Medicine, Washington University in Saint Louis

*Corresponding author:

Deshan Yang
Department of Radiation Oncology, Duke University
deshan.yang@duke.edu

## Abstract

**Background:** Quantification of cardiac motion on pre-treatment CT imaging for stereotactic arrhythmia radiotherapy patients is difficult due to the presence of image artifacts caused by metal leads of implantable cardioverter-defibrillators (ICDs). The CT scanners' onboard metal artifact reduction tool does not sufficiently reduce these artifacts. More advanced artifact reduction techniques require the raw CT projection data and thus are not applicable to already reconstructed CT images. New methods are needed to accurately reduce the metal artifacts in already reconstructed CTs to recover the otherwise lost anatomical information.

**Purpose**: To develop a methodology to automatically detect metal artifacts in cardiac CT scans and inpaint the affected volume with anatomically consistent structures and values.

**Methods**: Breath-hold ECG-gated 4DCT scans of 12 patients who underwent cardiac radiation therapy for treating ventricular tachycardia were collected. The metal artifacts in the images caused by the ICD leads were manually contoured. A 2D U-Net deep learning (DL) model was developed to segment the metal artifacts automatically using eight patients for training, two for validation, and two for testing. A dataset of 592 synthetic CTs was prepared by adding segmented metal artifacts from the patient 4DCT images to artifact-free cardiac CTs of 148 patients. A 3D image inpainting DL model was trained to refill the metal artifact portion in the synthetic images with realistic image contents that approached the ground truth artifact-free images. The trained inpainting model was evaluated by analyzing the automated segmentation results of the four heart chambers with and without artifacts on the synthetic dataset. Additionally, the raw cardiac patient images with metal artifacts were processed using the inpainting model and the results of metal artifact reduction were qualitatively inspected.

**Results**: The artifact detection model worked well and produced a Dice score of $0.958 \pm 0.008$. The inpainting model for synthesized cases was able to recreate images that were nearly identical to the ground truth with a structural similarity index of $0.988 \pm 0.012$. With the chamber segmentations on the artifact-free images as the reference, the average surface Dice scores improved from $0.684 \pm 0.247$ to $0.964 \pm 0.067$ and the Hausdorff distance reduced from $3.4 \pm 3.9$ mm to $0.7 \pm 0.7$ mm. The inpainting model's use on cardiac patient CTs was visually inspected and the artifact-inpainted images were visually plausible.

**Conclusion**: We successfully developed two deep models to detect and inpaint metal artifacts in cardiac CT images. These deep models are useful to improve the heart chamber segmentation and cardiac motion analysis in CT images corrupted by mental artifacts. The trained models and example data are available to the public through GitHub.

# 1  Introduction

Ventricular tachycardia (VT) results in over 300,000 cases of sudden cardiac death per year in the United States. Recent clinical studies[1-6] have shown stereotactic arrhythmia radiotherapy (STAR) as a viable option for the treatment of drug-resistant and recurrent VT [7] due to its ability to reach cardiac tissue beyond what is accessible with catheter ablation. However, the new STAR treatment requires that the cardiac motion of the patient is accounted for in treatment planning and radiation delivery. Due to setup uncertainty, cardiac motion, and respiratory motion, it is common for the prescribed target volume to be three times the volume of the arrhythmic tissue [7,8]. More precise motion management is necessary in clinical implementation of STAR [9,10].

Pre-treatment cardiac 4DCTs and respiratory 4DCTs could be used to inform the motion margins accurately for individual patients. However, over 90% of STAR patients have implantable cardioverter-defibrillators (ICDs) and the presence of the ICD leads in the cardiac area results in significant metal artifacts in the 4DCT scans. These metal artifacts prevent accurate registration, segmentation, and motion quantification of the heart and its chambers. Significant artifacts still exist as shown in the raw patient images in Figure 1 even after using the clinically available metal artifact reduction (MAR) techniques to reduce these artifacts. Advanced MAR techniques have been in development for decades beginning with projection data synthesis [11-14]. More MAR techniques continue to be developed in both pure projection and image domain, and a combination of the two especially exploring the uses of deep learning networks in recent years [15-19].

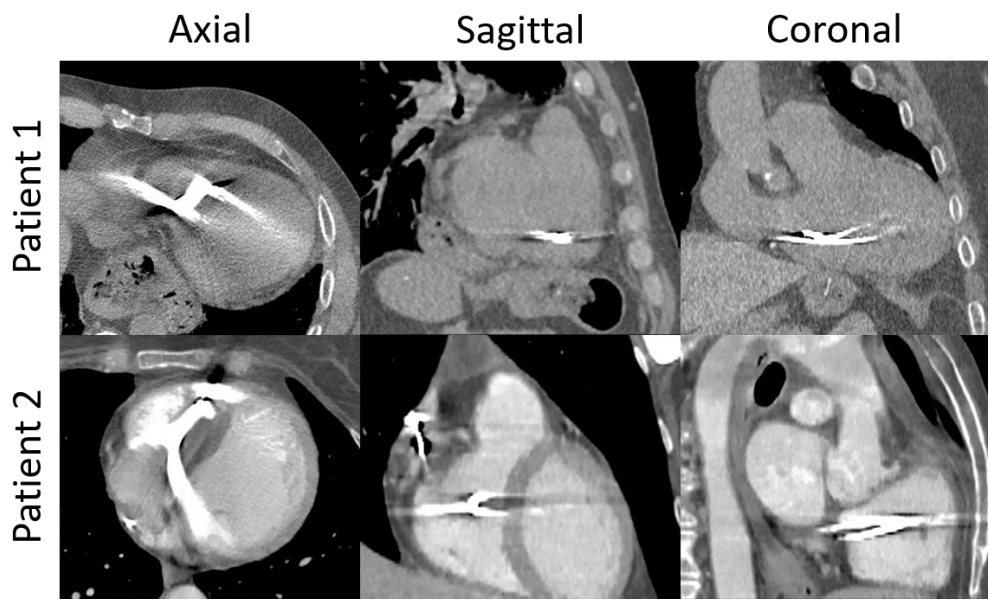

Figure 1: Examples of the metal artifacts present in two patient cases. There were varying degrees of artifacts throughout the images. The prevalence of the artifacts poses challenges for image analysis tasks, e.g. automatic segmentation of the cardiac chambers.

Initial MAR techniques were designed to tackle larger metal artifacts. There has been a significant push in recent years for smaller metal artifacts. The dyPAR+ method developed by Lossau et al in 2020 [20] focused on MAR in 4D-CT scans to reduce ICD lead artifacts while accounting for motion blurring. While dyPAR+ showed promising results, it required the CT projection data, and the corresponding ECG data recorded during the CT scan. This tool is not applicable to CT images that are already reconstructed. Additionally, Wang et al in 2017 [17] used a conditional GAN (cGAN) to reduce metal artifacts from CT images of patients who had cochlear implants. The cGAN used pre-cochlear implant artifact-free CT scans and the corresponding post-implant artifact-corrupted CT scans to train the cGAN model to reduce artifacts in new patients' post-implant CT scans. The network required a set of training images and known artifact-free ground truth, in line with the data available for this study.

In this study, we developed two separate deep networks for detecting and inpainting the ICD metal artifacts on patients' 4DCT images. The detection deep network was a 2D U-Net [21] model trained and tested on 120 manually contoured images. The artifact inpainting model was a GAN model trained on a digitally synthesized dataset that placed metal artifacts into cardiac CT images that previously did not have artifacts. Using the digitally synthesized dataset with corresponding ground truth images allowed the model to learn to use the image information surrounding the metal artifacts to infer the cardiac anatomy and fill the area corrupted by the metal artifacts with realistic and anatomically correct image intensity. This process of replacing the artifact mask with realistic values will be referred to as inpainting for the remainder of this paper. The corrected images can facilitate more consistent automated cardiac segmentation and image registration.

## 2 Material and methods
### 2.1 Artifact detection model
#### 2.1.1 Artifact detection materials

The data for training the artifact detection model were 12 patients' breath-hold ECG-gated 4DCT images with 10 cardiac phases per patient. These images were acquired with submillimeter axial resolution ranging from 0.48 x 0.48mm$^2$ to 0.69 x 0.69mm$^2$ and a consistent slice thickness of 1.5mm. The whole heart (the pericardial sac) and any metal artifacts caused by the ICD leads were manually contoured on each CT image. The segmented metal artifacts included high-intensity streaking and shadows in and near the heart including artifacts extending into the lung. Examples of the manual contours are shown in Figure 2. To be consistent, the manual contouring of the metal leads and artifacts began with simple thresholding of the cardiac region. From this initial thresholding, the artifacts were expanded into surrounding tissues where the primary bright spots and shadows could easily be identified. As the artifacts continued away from the source, the streaking artifacts became less condensed and were excluded from the artifact contour to preserve as much initial image information as possible. The artifacts were mainly within the heart but could extend beyond the cardiac area into the lungs and other surrounding organs. Artifacts that reached into these other organs were also segmented. The whole heart was segmented because we found during the preliminary study that the heart segmentation could be included in the loss function for deep model training to prevent the airways and bones from being detected as artifacts due to similar intensity values. Using the trained model to detect the artifacts only required the CT image. The model outputs were the segmentations of the whole heart and the metal artifacts.

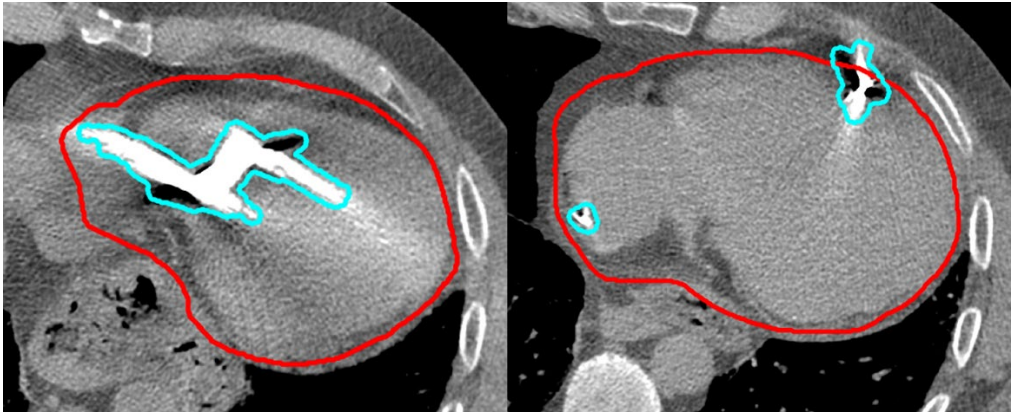

Figure 2: Demonstration of the whole heart and the metal artifacts manual segmentation. The whole heart was contoured in red, and the metal leads and artifacts were in cyan.

### 2.1.2 Artifact detection model design and training

The artifact detection model was built on the nnUNet v2 framework[21]. The model was designed to be a 2D model because the primary streaking artifacts existed mainly in the axial planes of the images. Two examples are shown in Figure 1. Using the 2D model design, by treating each image slice separately, also allowed for a greater number of training images. The detection network was trained with 2D slices from 80 3DCT images of eight patients for 400 epochs with 250 iterations per epoch. Twenty 3DCT images of two additional patients were used for validation during training. The training hyper-parameters were chosen to allow the model to stabilize on the provided validation dataset.

The model was trained using a five-fold cross validation on PyTorch 2.1.1, and NVIDIA GeForce RTX 3090 with CUDA toolkit 12.1. A complete model training for each fold took 4 hours, and inference from the trained model took three to four seconds per 3DCT.

### 2.1.3 Artifact detection model evaluation

The trained detection model was evaluated on images from all 10 phases of two patients set aside for testing. The model-predicted artifacts were compared to the manual segmentations and the Dice-Sorenson coefficients were calculated between the automated predictions and manual predictions.[22]. This comparison was useful for determining the tool's usefulness in a clinical setting using clinically relevant images. However, minor manual segmentation uncertainty might negatively affect the computed Dice scores. Additionally, we calculated the true positive rate (TPR) and true negative rate (TNR). The TPR was calculated after eroding the borders of the manual segmentation by three pixels to ensure that everything within the eroded segmentation was an artifact. The TNR was computed by measuring artifacts detected greater than 3 pixels away from the manually segmented artifacts.

## 2.2 Image inpainting model

The inpainting model's purpose was to automatically replace the detected artifact of the images with realistic image values. To train the inpainting model, we needed to use a dataset that differed from what was used for training the artifact detection model because the inpainting model needed ground truth images. In addition, this model was built separately from the artifact detection model so that it could take any artifact mask, e.g. manually contoured artifacts, and was not confined to metal artifacts detected by the first detection model.

### 2.2.1 Artificial synthesis of the artifact dataset synthesis

To create a dataset to train the inpainting model, we used the 120 manually contoured patient CT images explained in Section 2.1.1 and 148 of the CT images from the TotalSegmentator [23] training dataset. The 148 cases were manually selected to contain the entirety of the heart without pre-existing metal artifacts.

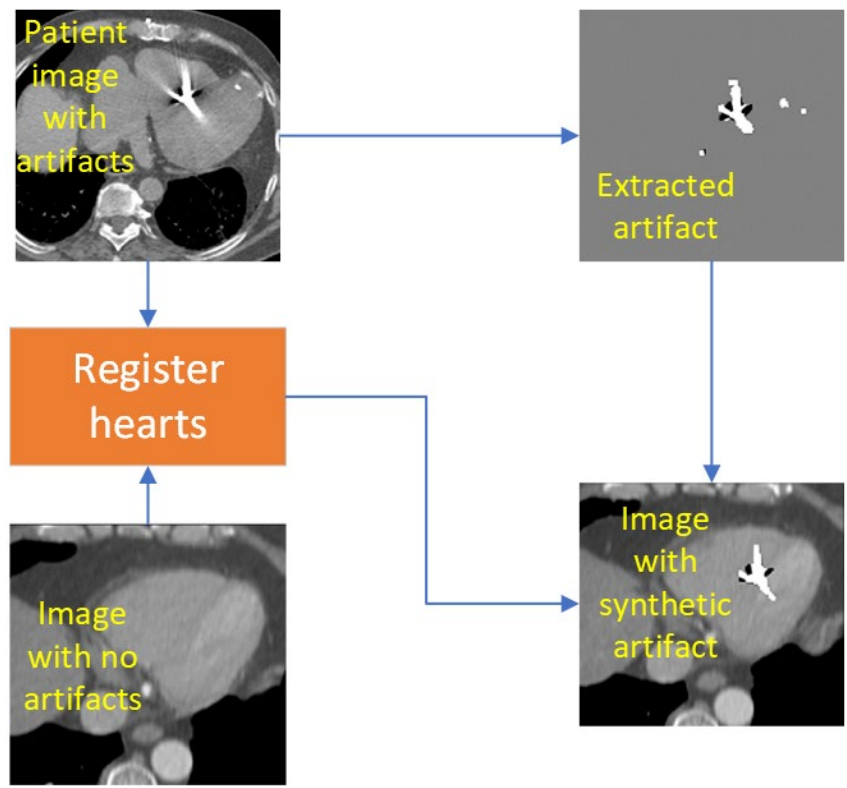

Figure 3: Showing the creation of the synthetic artifact dataset for training of the inpainting model.

Placing the segmented metal artifacts from the STAR patient images into the artifact-free cardiac CTs resulted in images with known artifacts obscuring the underlying ground truth images. The procedure is illustrated in the workflow shown in Figure 3. We first registered the images by aligning the centers of the hearts in both datasets. Once the hearts were aligned, any contoured metal artifacts from the patient dataset that laid within 3 pixels (4.5 mm) of the heart segmentation were placed into the artifact-free CT image. The 3-pixel border surrounding the heart was selected due to noticing metal artifacts on the border of the heart and streaking into the lungs and surrounding organs on many of the STAR patients' images. By including this margin, the inpainting model learned not only to correct the artifacts within the pericardium, but also learned how to predict the structural organ borders using surrounding image information.

### 2.2.2 Inpainting model design

The inpainting model was designed based on the Pix2Pix [24] framework which is a GAN model converting one image to another. The goal of the inpainting model was to replace the artifact portion of the image with a realistic image voxel intensity based on image voxels surrounding the artifacts. The model accomplished this goal by 1) predicting the filled image and 2) using a discriminator network to ensure the artificially filled image was indistinguishable from the ground truth image. The network architecture chosen for this task included four encoder layers and four decoder layers leading to 4.8 million learnables.

To train the model, we first extracted image patches from the artificially placed artifact images described in section 2.1.2. These patches were 80x80x32 pixels with the axial patch dimensions chosen to be larger than the largest of the artifacts on any given slice across all images. This allowed for the model to utilize the image information surrounding the detected artifacts to inform the inpainting prediction. The 32-slice patch thickness was determined through preliminary studies to balance the GPU memory usage.

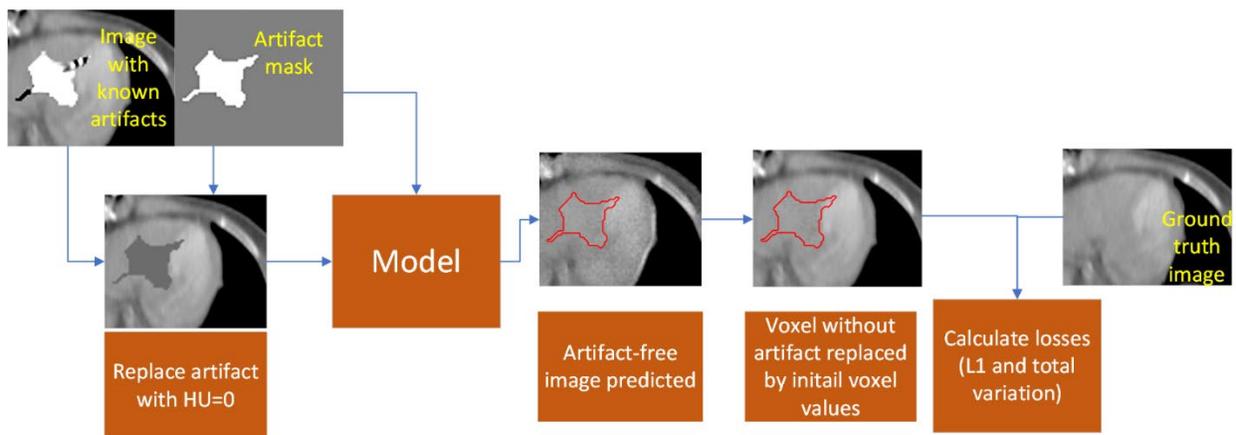

Figure 4: Workflow of inpainting model design and loss calculations. L1 loss and total variation loss are calculated only within the artifact since all other voxels will be identical to the ground truth.

The model workflow is shown in Figure 4. We used two input channels. The first channel was the image with artifacts after the artifact voxel were replaced with an HU of 0. Overwriting the artifact to voxel values closer to the expected intensity values allowed the model to correct the artifacts more easily. The second channel was the binary mask of the artifact. From this image with the artifact set to an HU of 0, the model predicted the final artifact-free image. To retain the artifact-free portion of the original image, we replaced the predicted image outside of the artifact with the corresponding voxel intensities from the initial image since that remained unaffected by the metal artifacts.

The inpainting model was trained by minimizing the generator loss, discriminator loss, L1 loss, and Total Variation (TV) loss between the prediction results and the ground truth image[25]. The L1 loss was calculated only on the voxel values in the logical artifact mask. The TV loss was calculated on the difference image, which was calculated by subtracting the model prediction from the ground truth image. The TV loss ensured that the inpainted artifact region transited smoothly to the original image outside the artifact region. The TV loss was weighted half as strongly as the L1 loss.

### 2.2.3 Evaluation of the inpainting model

To evaluate the inpainting model, it is important to look not only at how well it predicts the true underlying pixel value but also at how much of the anatomical structure information is recovered. Therefore, in addition to looking at the structural similarity metric of the inpainted sections of the image, we also compared the automatic segmentation of the heart chambers on the inpainted image to the known segmentation ground truth in the original image without artifacts. The evaluation procedure is shown in Figure 5. Because all image voxels outside the metal artifacts exactly matched the ground truth image by design, the comparison metrics were only calculated within the artifact region. In addition to the whole structure Dice score, we also used the surface Dice score [26] because the surface Dice score correlated better with the structural surface position. The tolerance value used for the surface dice calculation was three pixels. The automated segmentation tool used for these calculations was TotalSegmentator[23]. To detect the model's reliance on training images vs validation images, a five-fold cross validation was performed with no significant impact on the model's performance.

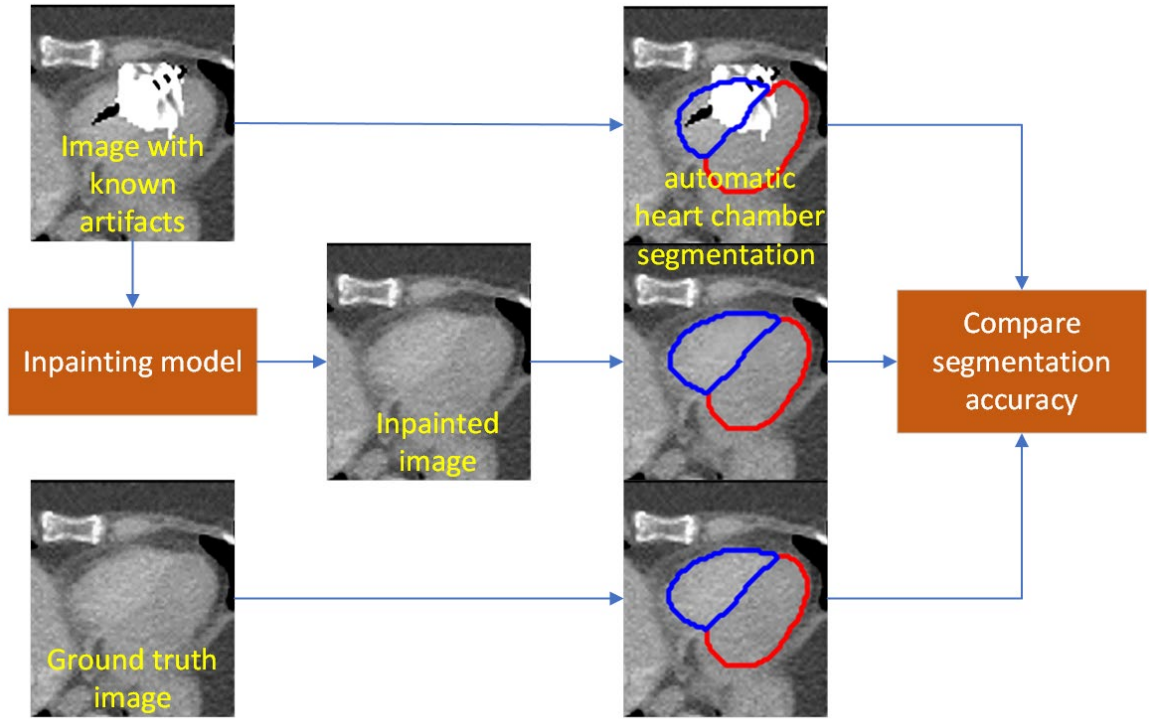

Figure 5 The procedure to evaluate the inpainting model. To ensure fairness in the testing, the segmentation was all performed by the same automatic segmentation tool.

## 3 Results
### 3.1 Artifact detection results

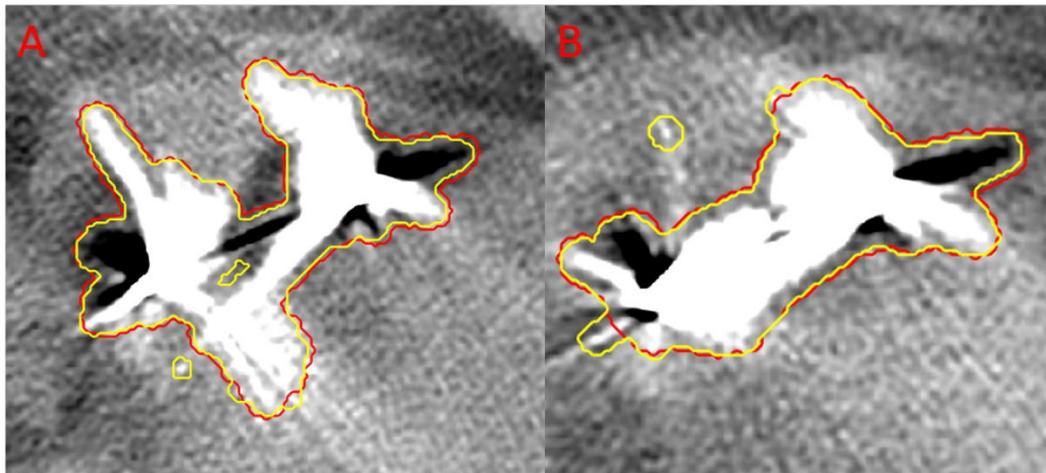

Figure 6: Demonstration of the automatically predicted artifact mask for one test case. The yellow outline was the predicted artifact mask. The red was the manual segmentation of the artifacts.

The artifact detection model was tested on 20 3DCT volumes from two additional patients with manual contours. As shown in Figure 6, the model performance on patient images was visually excellent. The primary disagreements were at the edges of the segmentations, caused by minor manual segmentation inconsistencies. For example, in image A, there was a hole in the center of the automatically detected artifact, while images A and B included identified minor artifacts outside of the manual segmentation. Table 1 shows the quantitative evaluation results for the artifact detection model.

Table 1: The quantitative performance of the artifact detection model on the two test patients. The methodology for the true positive and true negative rates is explained in Section 2.1.2.

| Metric | Results |
|---|---|
| Dice score | $0.958 \pm 0.008$ |
| True Positive rate | $0.983 \pm 0.007$ |
| True Negative rate | $0.972 \pm 0.005$ |

## 3.2 Inpainting model results
### 3.2.1 Segmentation of inpainted results

We computed the structural similarity metric between the inpainted and ground truth images for the test cases and achieved a score of $0.988 \pm 0.012$ across 148 cases that were left out of the model training. The results indicate that the model was successful in utilizing the surrounding tissue information to correctly predict what image intensity would accurately fill in the affected volume.

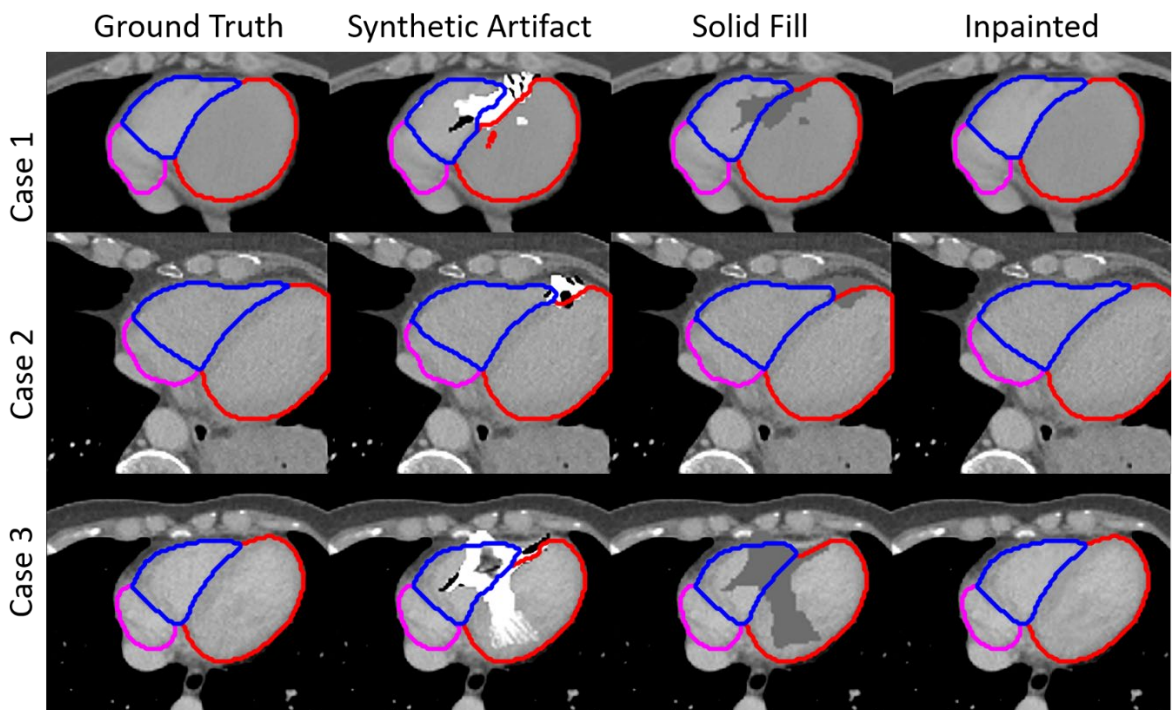

Figure 7: The automatic segmentation results for three cases. In the "Solid Fill" column, the artifact voxels were overwritten by an HU of 0.

Figure 7 shows the resulting automatic segmentation of three test cases following the inpainting model. The second column shows the artifact placed in the artifact-free image in a way that was consistent with what we observed in patients' 4DCT images shown in Figure 1. The third column shows that overwriting the artifact region with an HU value of 0 did a decent job of reducing the effect of the artifacts on the segmentation results. However, the auto-segmentation tool still struggled to find borders in the solid fill images as seen in cases 1 and 2. The segmentation results on the inpainted images shown in the fourth column were nearly identical to the ground truth images upon visual inspection of the automated segmentation results.

A quantitative comparison of the auto-segmentation accuracy of the cardiac chambers is shown in Table 2. The average surface Dice score, whole structure Dice score, mean surface-to-surface distance, max Hausdorff distance, and 95% Hausdorff distance of each structure were calculated and averaged across all 148 test cases. These metrics were calculated for heart chambers where the artifacts existed in the images with artifacts added, with the artifacts replaced with HU of 0, and with the artifacts inpainted. Artifacts were not present in all chambers for each test patient and those chambers were left out of the analysis. There were 409 chambers in total affected by artifacts in the test cases. The comparison showed that while a solid fill of the artifacts with 0 HU improved the auto-segmentation, the inpainting results easily outperformed the solid filling. The greatest improvement was seen in the surface-to-surface distance measurements. The inpainting method's ability to use surrounding tissue information to predict anatomical borders resulted in much greater accuracy of segmentations.

Table 2: Showing the improvement in cardiac chamber segmentation accuracy for the inpainted image compared to the initial artifact and simply filling the artifact volume with an HU of 0.

|  | Surface Dice | Dice | Mean surface-to-surface distance | Hausdorff distance | 95% Hausdorff distance |
|---|---|---|---|---|---|
| Artifact Uncorrected | 0.684 ± 0.247 | 0.931 ± 0.120 | 0.9 ± 1.6 mm | 3.4 ± 3.9 mm | 2.9 ± 3.6 mm |
| Artifact filled with HU=0 | 0.765 ± 0.222 | 0.956 ± 0.107 | 0.4 ± 0.4 mm | 1.7 ± 1.5 mm | 1.4 ± 1.1 mm |
| **Artifact Inpainted** | **0.964 ± 0.067** | **0.995 ± 0.020** | **0.1 ± 0.1 mm** | **0.7 ± 0.7 mm** | **0.3 ± 0.5 mm** |

We also tested the inpainting model on the STAR patients' 4DCT images. Example results are shown in Figure 8. We only qualitatively evaluated the inpainting results visually due to not having the unaffected ground truth images. The results showed that the inpainting model could successfully extrapolate the borders of the cardiac substructures.

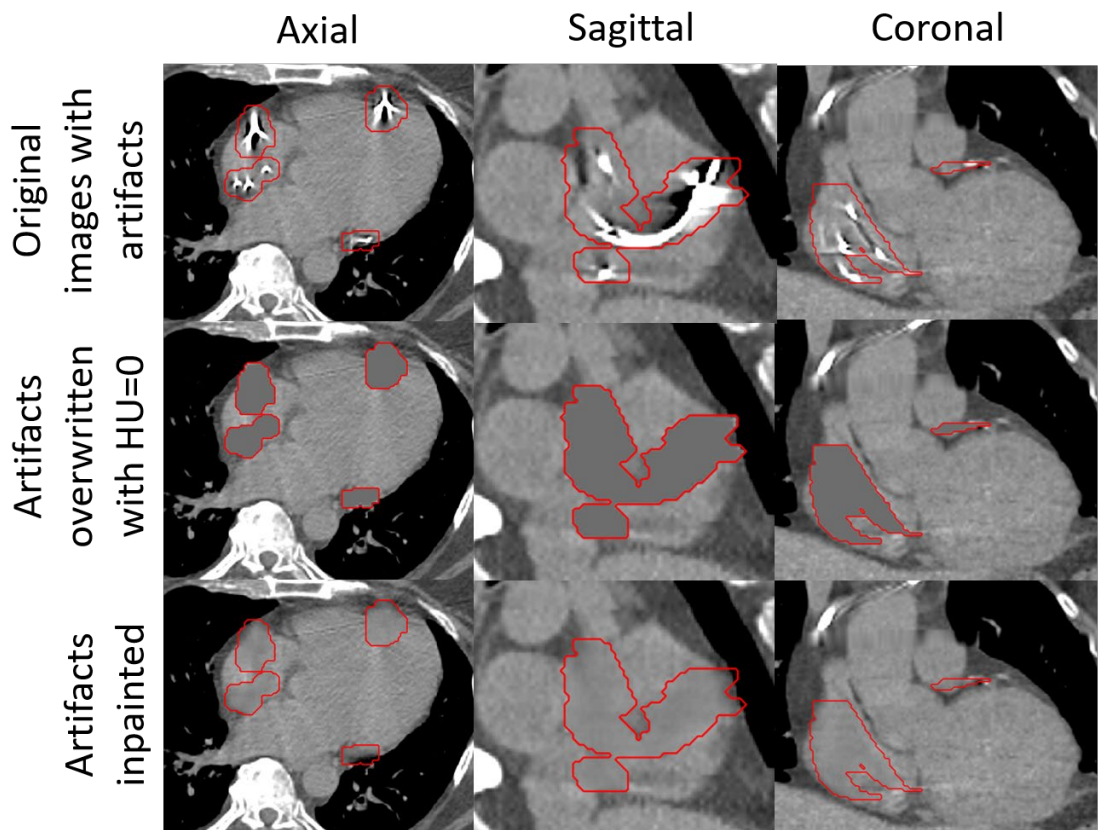

Figure 8: The inpainting results on patients' 4DCT images showing anatomically realistic inpainting of artifacts.

## 4 Discussion

While the proposed inpainting method resulted in visually plausible replacements of the metal artifacts, quantification of the results remains challenging. An automatic 3rd party tool such as TotalSegmentator was used in this study to segment the heart chambers affected by the metal artifacts. However, the comparison of the auto-segmentation results was still reliant on the tool and thus might be biased. Replacing auto-segmentation with manual segmentation would also be confounding due to potential inter-observer variations.

The artifact placement process used to create the artificial dataset was designed to ease the development of the inpainting models rather than to provide the most realistic artifacts. In future work, more accurate artificial dataset creation methods can be used such as inserting the metal artifacts into the volume and simulating the CT results as mentioned in [19,27]. This may result in more consistent artifacts in the training images and may improve the inpainting model further.

Another concern with the inpainting model is the smoothness of the inpainting results compared to the noise in the original images. While the smoothness in the images generated by the inpainting model has minimal effects on radiation dose calculation or image segmentation, the inpainted images look artificial, as shown in Figure 8. There are ways to improve the looks of the inpainting results. For example, the image noise texture can be extracted from the original image and then added to the inpainted image to make the inpainted section look more realistic.

The use of inpainting models for reducing metal artifacts has shown promise as seen in Wang et al. 2018 [17] but remains a rather underutilized methodology. While this work and the work seen in Wang et al. focused on using GAN models, there is exciting work being done with other deep learning networks such as diffusion models that may provide more realistic results[28]. However, deep diffusion models are still not a fully mature methodology and would require much greater computation power than simpler GAN models.

# 5 Conclusion

Two deep networks were developed to identify and correct the metal artifacts in cardiac CT images. The deep models provided realistic inpainting on artifact-corrupted patient images and allowed for the recovery of anatomical borders which could facilitate more consistent heart substructure segmentation on STAR patients' pre-treatment 4DCTs.

# 6 Dissemination

The trained models used for this work can be located at the following GitHub directories; https://github.com/deshanyang/ICD-Artifact-Detection and https://github.com/deshanyang/Cardiac-Inpainting.

## Acknowledgment

This research was supported by the National Heart, Lung, and Blood Institute (NHLBI) grant R01-HL148210, and the National Institute of Biomedical Imaging and Bioengineering (NIBIB) grant R01-EB029431.